\documentclass[twocolumn,english,aps,prb,showpacs,byrevtex,amsmath,amssymb,superscriptaddress]{revtex4-1}
\usepackage{graphicx}
\usepackage{epstopdf}
\usepackage{dcolumn}
\usepackage{bm}
\usepackage{upgreek}
\usepackage{graphicx}
\usepackage{epstopdf}
\usepackage{dcolumn}
\usepackage{bm}
\usepackage{gensymb}
\usepackage{upgreek}
\usepackage{booktabs}
\usepackage{hyperref}

\usepackage{amssymb,mathtools}
\usepackage{color}
\usepackage{amsmath} 
\usepackage[makeroom]{cancel}

\usepackage{tikz,xcolor,hyperref}
\definecolor{lime}{HTML}{A6CE39}
\DeclareRobustCommand{\orcidicon}{%
	\begin{tikzpicture}
	\draw[lime, fill=lime] (0,0)
	circle [radius=0.16]
	node[white] {{\fontfamily{qag}\selectfont \tiny ID}};
	\draw[white, fill=white] (-0.0625,0.095)
	circle [radius=0.007];
	\end{tikzpicture}
	\hspace{-2mm}
}

\foreach \x in {A, ..., Z}{%
	\expandafter\xdef\csname orcid\x\endcsname{\noexpand\href{https://orcid.org/\csname orcidauthor\x\endcsname}{\noexpand\orcidicon}}
}

\begin{document}

\title{Coexistence of Rashba and Dirac dispersions on the surface of centrosymmetric topological insulator decorated with transition metals}

 \author{Giuseppe Cuono\orcidB}
\affiliation{Consiglio Nazionale delle Ricerche (CNR-SPIN), Unità di Ricerca presso Terzi c/o Università “G. D’Annunzio”, 66100, Chieti, Italy}

\author{Rajibul Islam\orcidC}
\affiliation{Department of Physics, University of Alabama at Birmingham, Birmingham, AL, USA}

\author{Amar Fakhredine\orcidF}
\affiliation{Institute of Physics, Polish Academy of Sciences, Aleja Lotnik\'ow 32/46, 02668 Warsaw, Poland}

\author{Carmine Autieri\orcidA}
\email{autieri@magtop.ifpan.edu.pl}
\affiliation{International Research Centre Magtop, Institute of Physics, Polish Academy of Sciences, Aleja Lotnik\'ow 32/46, PL-02668 Warsaw, Poland}
\affiliation{SPIN-CNR, UOS Salerno, IT-84084 Fisciano (SA), Italy}

\date{\today}
\begin{abstract}
The Dirac cone originates from the bulk topology, yet its primary contribution comes from the surface since spatially, the Dirac state emerges at the boundary between the trivial and topological phases. At the same time, the Rashba states emerge in regions where inversion symmetry is broken. On the surface of the centrosymmetric topological insulators, both Rashba and Dirac bands are present and their hybridization produces the giant Rashba effect, modifying both Rashba's and Dirac's bands. Therefore, pure Rashba and Dirac fermions are inherently incompatible on the surface of centrosymmetric topological insulators if the material is homogeneous. Inspired by recent experiments, we focused on the (111) polar surface of PbSe, which becomes a topological crystalline insulator under compressive strain, and we established the conditions under which a topological system can simultaneously host pure Rashba and Dirac surface states close to the Fermi level.
The coexistence of pure Dirac and Rashba dispersions is only possible in a non-homogeneous centrosymmetric topological insulator, where the spatial origins of the two bands are effectively separated.
In the experimentally observed case of PbSe, we demonstrate that a metallic overlayer induces a strong electrostatic potential gradient in the subsurface region, which in turn generates the electric field responsible for Rashba splitting in the subsurface layers. Consequently, in PbSe(111), the Rashba states arise from subsurface layers, while the Dirac states live mainly on the surface layers. Finally, we compare the properties of the Rashba in the trivial and topological phases; the calculated Rashba coefficient agrees qualitatively with the experimental results.
\end{abstract}

\maketitle

\section{Introduction}
Topological insulators were discovered as a new phase of matter exhibiting insulating behavior in their bulk while hosting conductive surface states protected by time-reversal symmetry. These surface states are characterized by a Dirac cone, where the linear energy-momentum relationship resembles that of relativistic particles\cite{RevModPhys.82.3045,Brzezicki_2020}. The coverage of the surface (also called decoration) has been used to tune topological properties and induce topological Lifshitz transitions\cite{Yang2019,PhysRevB.105.235304}.
Topological crystalline insulators have topological surface states protected by the combination of time-reversal and crystalline symmetries. 
While the topological insulators host a Dirac point at $\overline{\Gamma}$ (the $\Gamma$ point projected on a given surface), the topological crystalline insulators host a Dirac point in a different position of the Brillouin zone. The discovery of the topological crystalline insulating (TCI) phase in SnTe\cite{Fu11,Hsieh12,Lau19} and some of its substitutional alloys, such as Pb$_{1-x}$Sn$_x$Te\cite{Xu12} and Pb$_{1-x}$Sn$_x$Se\cite{Dziawa12,Islam19,Cuono2022}, has sparked significant research interest in exploring this class of materials in bulk, at interfaces, and on surfaces\cite{PhysRevB.100.041408,PhysRevB.103.245307}. It has been found that the key properties of these materials vary with size or dimensionality (2D or 3D phases), meaning that their characteristics differ when transitioning from the bulk \cite{Plekhanov14,Wang20,Barone13} to thin films \cite{Slawinska20,Volobuev17,Liu18}. Several works investigate the nanowires of topological crystalline insulators as candidates for devices to generate Majorana fermions\cite{PhysRevB.105.075310,kawala2024topologicalproperties110snte,D4NH00019F,doi:10.1021/acsanm.4c00506}. 

The Rashba spin splitting (RSS) is a relativistic spin-splitting that can arise in the absence of inversion symmetry. Analogously, the Dirac cone also produces Dirac spin splitting (DSS). The RSS was deeply studied in several classes of materials\cite{10.1063/1.5137753,PhysRevB.100.245115,HUSSAIN2022169897,PhysRevB.111.L041202,PhysRevB.108.035121}, and it has relevant consequences on the properties of materials as bilinear magnetoresistance\cite{PhysRevB.106.L241401}, persistent spin-helix\cite{PhysRevMaterials.3.084416}, spin-to-charge conversion\cite{PhysRevB.107.165140} and high-performance devices enabling electrical control of the spin information\cite{https://doi.org/10.1002/adma.202002117}.
The interplay between Dirac points and Rashba effects has also been studied in centrosymmetric topological materials, giving rise to the giant Rashba (also called hybridized Rashba) \cite{https://doi.org/10.1002/adma.201604185,https://doi.org/10.1002/adfm.202008885,PhysRevLett.107.186405,doi:10.1021/acs.nanolett.4c03802} and in 2D noncentrosymmetric systems 
\cite{Xue2024,doi:10.1021/acsami.8b18582,MortezaeiNobahari2024,PhysRevB.111.075407}. In both previous cases, the surface states result from a hybridization of Rashba and Dirac fermions, but without the coexistence of pure Rashba and Dirac fermions. In Bi$_2$Te$_3$-based topological materials, the Rashba was engineered by termination, but the Dirac and Rashba states did not hybridize because they were well separated in energy\cite{PhysRevB.110.205113}.
To our knowledge, the coexistence of pure Rashba and Dirac fermions at the same energy was predicted up to now only in 3D ferroelectric topological materials\cite{PhysRevLett.117.076401,PhysRevB.107.075143} or topological systems without inversion symmetry\cite{PhysRevB.104.085113,PhysRevResearch.4.023114} since the Rashba states rise from the bulk and the Dirac states from the surface.

\begin{figure}[t!]
\centering
\includegraphics[width=\columnwidth,angle=0]{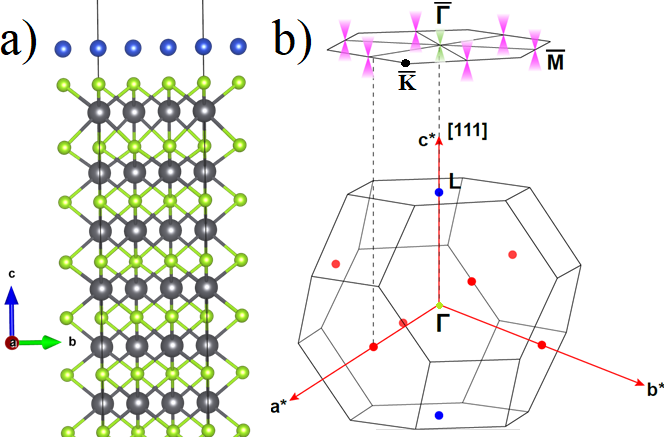}
\caption{(a) Crystal structure of the TCI grown along the (111) direction with the Se termination. The TCI is covered by a single layer with the 3$d$ transition metals on top of Se atoms. The black, green and blue atoms represent the Pb, Se and Cu atoms, respectively. a, b and c are the real-space lattice vectors. (b) Three-dimensional Brillouin zone of the bulk TCI with three vectors of reciprocal lattice a*, b* and c*. In red and blue, there are the 6 and 2 L-points of the three-dimensional Brillouin zone, respectively.  
The projections of the reciprocal space on the (111) surface of the TCI with the surface Dirac cones at $\overline{\Gamma}$ (plotted in green) and $\overline{M}$ (plotted in purple) are shown.}
\label{structure}
\end{figure}

A recent angle-resolved photoemission spectroscopy (ARPES) investigation demonstrated that growing metal epilayers on the (111) orientation of topological Pb$_{1-x}$Sn$_x$Se can create the coexistence of pure Rashba and Dirac fermions on the surface\cite{phdthesisBartek,TUROWSKI2023155434}. 
The experiment consists of the MBE-grown (111) Pb$_{1-x}$Sn$_x$Se studied in the topological phase and under the influence of the deposition of 3$d$ transition metal elements on the surface. Due to the low dimensionality and the minimal amount of 3d elements, no long-range magnetism was observed. The conduction band of the (111) Pb$_{1-x}$Sn$_x$Se near the $\overline{\Gamma}$ point after deposition of the 3$d$ coverage on the sample surface undergoes splitting in momentum space due to the Rashba effect. This Rashba splitting has been previously reported in other TCI systems\cite{https://doi.org/10.1002/adma.201604185,https://doi.org/10.1002/adfm.202008885}, but not in coexistence with a pure Dirac point. The experimental data are described with a fitting procedure that does not take into account the hybridization of the DSS and RSS. Additionally, well-resolved quantum levels (QL) from the 2D electron gas appear after the deposition of the metal and become increasingly visible with the amount of the 3$d$ element on the surface of the sample. The 2D electron gas state undergoes a splitting in momentum, giving rise to bands visible between DSS and RSS. Experimentally, the Rashba coefficient increases linearly for low coverage and saturates at high coverage. 

The surface bands cannot simultaneously exhibit both Dirac and Rashba characteristics\cite{https://doi.org/10.1002/adfm.202008885}, necessitating further investigation to resolve this apparent contradiction. In this paper, we solve this puzzle by demonstrating that the metal decoration produces a strong gradient of the electrostatic potential. The gradient produces an electric field that pushes the origin of the Rashba band into the subsurface, decoupling the Rashba from the Dirac bands. Using strained PbSe over SnTe enhances the Rashba coefficient in the conduction band, due to the larger spin-orbit coupling (SOC) of Pb compared to Sn.  Our theoretical investigation will focus on the regime of a large surface covered where the Rashba coefficient saturates.
The paper is divided into three Sections: while in the second Section we present the main results, the last Section is devoted to the conclusions.

\begin{figure}[t!]
\centering
\includegraphics[width=\columnwidth,angle=0]{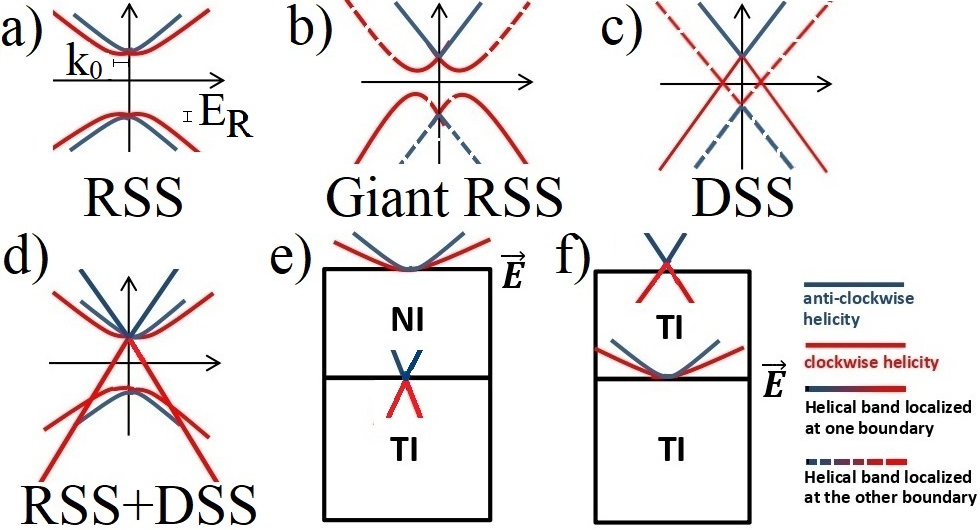}
\caption{RSS and DSS of an asymmetric slab in the limit of large thickness. a) Pure RSS on the surface of normal insulators (NI). b) Interplay between RSS and DSS on the surface of topological insulators. c) Pure DSS on the surface of topological insulators with negligible spin-orbit or in which the RSS is absent. d) Coexistence of pure Rashba and Dirac fermions in the surface band structure of the TCI. The Dirac point was put close to the valence band, as found in the experimental results.  
Possible options to have the coexistence of pure Rashba and Dirac fermions: e) the surface becomes trivial and the Dirac bands lie in the subsurface and f) an electric field ($\vec{E}$) in the bulk pushes the Rashba states to lie in the subsurface. The Rashba coefficient is defined as the ratio $\alpha_R$=$\frac{2E_R}{k_0}$ where E$_R$ and k$_0$ are represented in panel a) and described in the text.}
\label{coexistence}
\end{figure}

\begin{figure*}[t!]
\centering
\includegraphics[height=\columnwidth, angle=270]{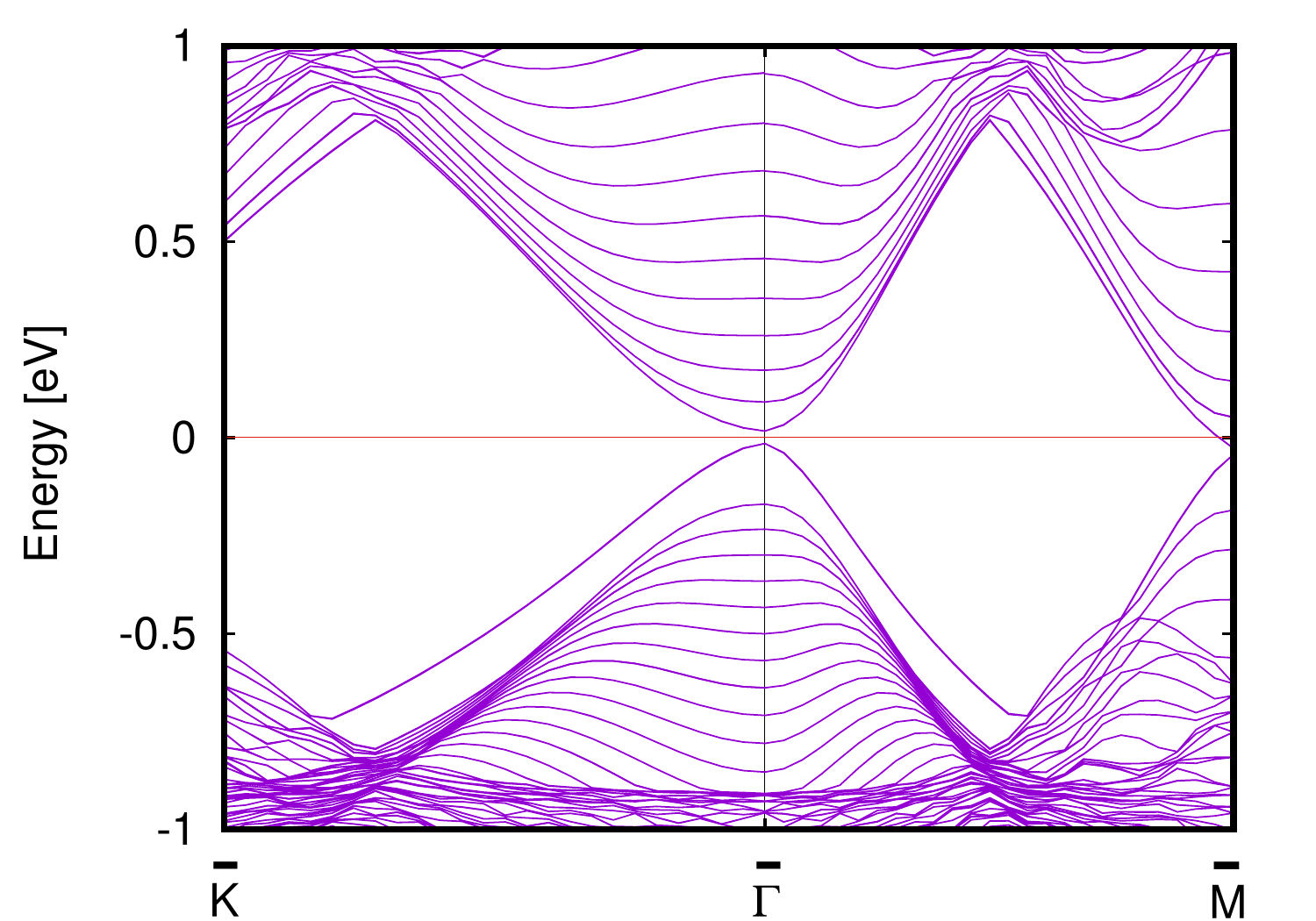}
\includegraphics[height=\columnwidth, 
angle=270]{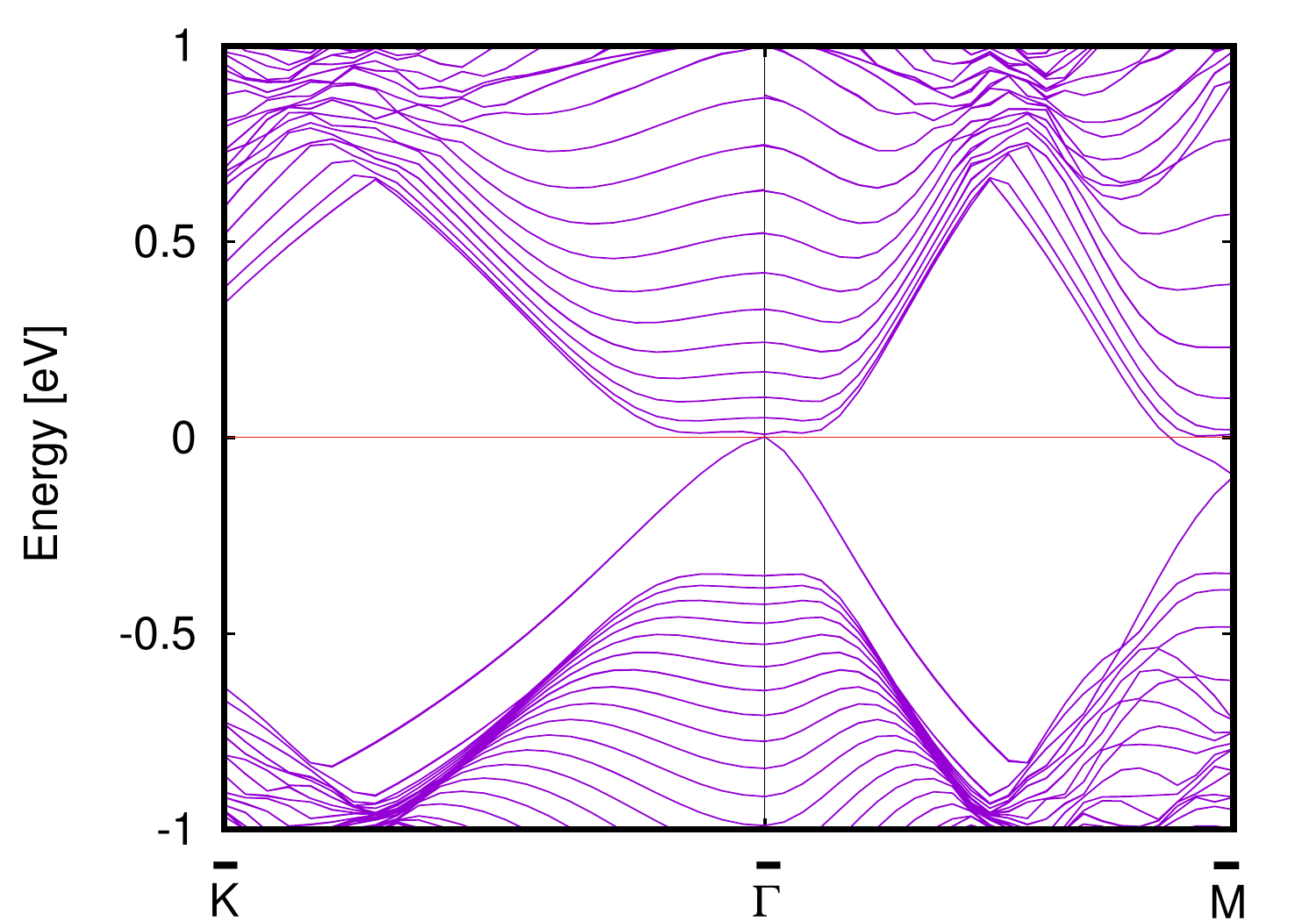}
\caption{Band structure of a symmetric uncovered slab in the trivial phase (left panel). Band structure of a symmetric uncovered slab in the topological phase (right panel). The Fermi level is set at zero energy.
}
\label{symmetric}
\end{figure*}

\section{Results}

In the next subsections, we define the problem and discuss the slab with the uncovered surface. Then, we move to the study of the slabs covered with 3$d$ metal, where we calculate the electronic structure, the surface-resolved band structure, the Rashba coefficient, and the electrostatic potential. 

\subsection{Difficulties for the coexistence of pure Rashba and Dirac fermions}

We investigated the interplay between Rashba and Dirac on the (111) surface of the TCIs when they are decorated with transition metals. The crystal structure is shown in Fig. \ref{structure}(a). 
The system consists of atomic layers with opposite charges; therefore, the surface is polar. 
When we consider the (111) surface of the TCI, the system hosts a Dirac cone at $\overline{\Gamma}$ and 3 Dirac cones at the $\overline{M}$ points.\cite{PhysRevB.88.241303,kazakov2025topologicalphasediagramquantum} The 2D Brillouin zone is reported in Fig. \ref{structure}(b). At the corners of the hexagonal Brillouin zone, we have the points $\overline{K}$ that do not host surface Dirac points but will be considered in the high-symmetry k-path for the band structure.  
The Dirac points correspond to the projection of the bulk L points onto inequivalent time-reversal invariant momenta at $\overline{\Gamma}$ and each of $\overline{M}$ high-symmetry surface points of the BZ projected on the (111) surface. The band structure consists of a Dirac cone with topological surface states crossing at the Dirac point. A schematic of the Fermi surface of the (111) surface shows the positions of all Dirac cones in high-symmetry surface points and the (110) mirror plane, responsible for mirror-symmetry protection of the surface states.\cite{Dziawa12,Brzezicki_2020}

In the case of a symmetric slab, the band structure is twofold degenerate at every $k$-point, including the Dirac cone and the system does not host RSS. 
The Rashba band structure of an asymmetric trivial insulator slab is instead shown in Fig. \ref{coexistence}(a).
The two Dirac points of an asymmetric topological insulator (where the Rashba is neglected) are shown in Fig. \ref{coexistence}(c). The two Dirac points correspond to one for each surface. The interplay between them generates a giant Rashba (shown in Fig. \ref{coexistence}(b)), but pure Dirac fermions are not present anymore. 
The giant Rashba effect comes from a strong interplay between topology and Rashba that is rarely observed in other topological systems\cite{https://doi.org/10.1002/adfm.202008885}, 
and it is characterized by a large value of the Rashba coefficient $\alpha_R$, where $\alpha_R$=$\frac{2E_R}{k_0}$, as shown in Fig. \ref{coexistence}(a). E$_R$ is the energy difference between the crossing point and the minimum of the lower branch, while k$_0$ is the distance in the k-space between the crossing point and the minimum of the lower branch.
What was observed in the recent experiments\cite{phdthesisBartek} is the coexistence of pure Rashba and Dirac fermions as shown in Fig. \ref{coexistence}(d). The task of the present paper is to describe the conditions that allow this coexistence in the electronic structure.
Two distinct options are possible to obtain this exotic band structure. To avoid strong hybridization between DSS and RSS, one of the Dirac bands and the Rashba bands needs to be shifted in the inner layers of the materials. They are plotted in Fig. \ref{coexistence}(e) and  Fig. \ref{coexistence}(f). While for the band structure, both options are equivalent, they are physically distinct. We prove in the rest of the paper that the coverage of the 3$d$ element on PbSe(111) pushes the Rashba bands in the inner layers, leaving the unperturbed DSS on the surface.

\subsection{Slabs with uncovered surface}

In this subsection, we show the band structures of PbSe(111) calculated for symmetric and asymmetric uncovered slabs in their trivial and topological phases.
The transition from the topological to the trivial phase can be obtained by strain, as described in Appendix A.

In the case of a symmetric slab with Pb-terminated surfaces, the band structure is twofold degenerate at every $k$-point and does not present RSS. 
In the trivial phase, we have a direct band gap at the $\bar{\Gamma}$ point, as shown in the left panel of Fig. \ref{symmetric}.
In the topological phase, a double-degenerate Dirac cone is present at the $\bar{\Gamma}$ point, as shown in the right panel of Fig. \ref{symmetric}. The energy of the Dirac point is very close to the conduction band in the Pb-terminated symmetric case. The pure Dirac point exists as long as the thin film is thick enough to avoid hybridization between the two surfaces.

\begin{figure*}[t!]
\centering
\includegraphics[height=\columnwidth, angle=270]{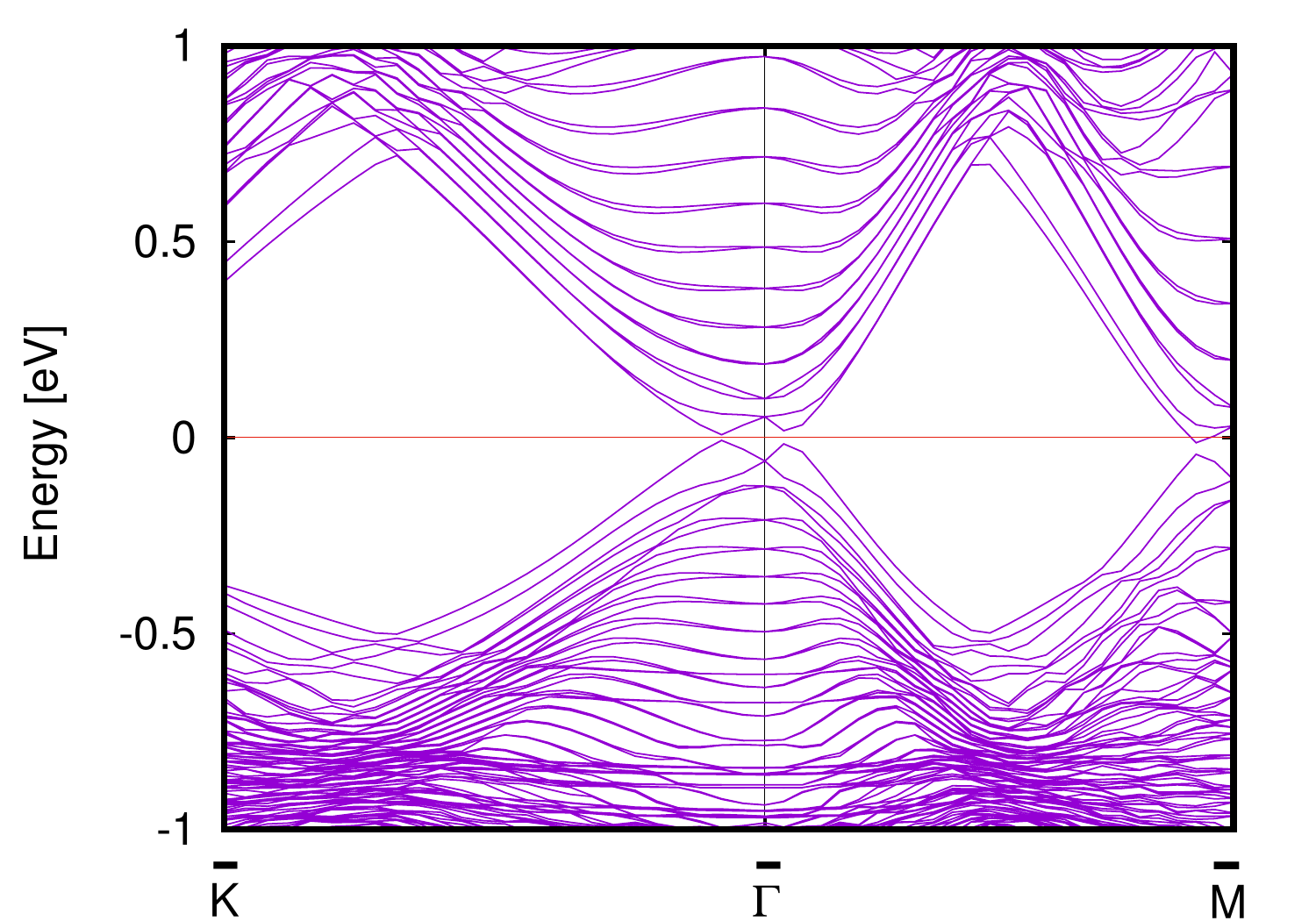}
\includegraphics[height=\columnwidth, angle=270]{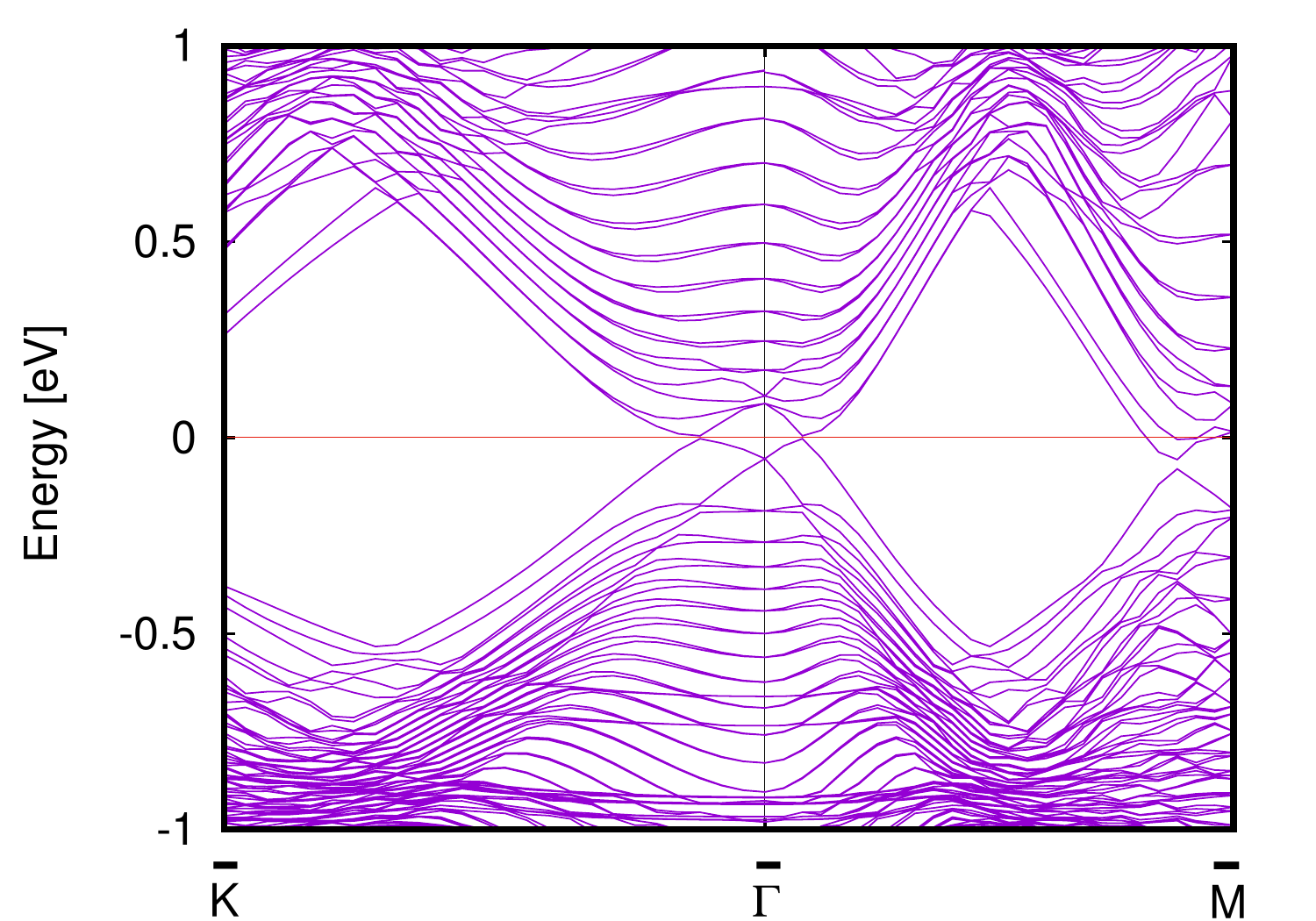}
\caption{Band structure of an asymmetric uncovered slab in the trivial phase (left panel). Band structure of an asymmetric uncovered slab in the topological phase (right panel). The Fermi level is set at zero energy.
}
\label{asymmetric}
\end{figure*}

\begin{figure}[t!]
\centering
\includegraphics[height=\columnwidth, angle=270]{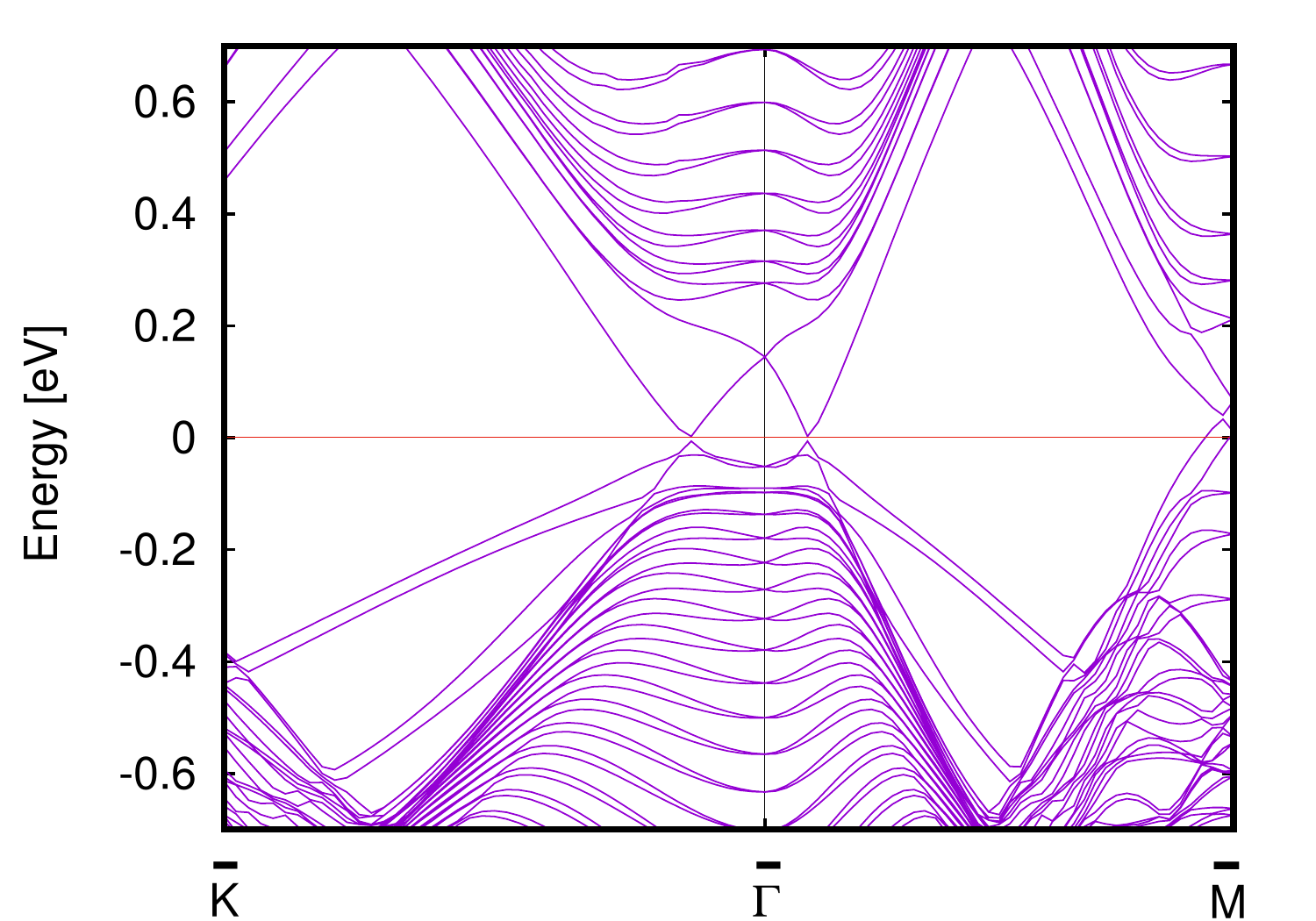}
\caption{Band structure of an asymmetric slab covered by the 3$d$ metal in the hollow position and the topological phase. The Fermi level is set at zero energy.
}
\label{Bandstructure_3d_hollow}
\end{figure}

\begin{figure}[t!]
\centering
\includegraphics[width=\columnwidth,angle=0]{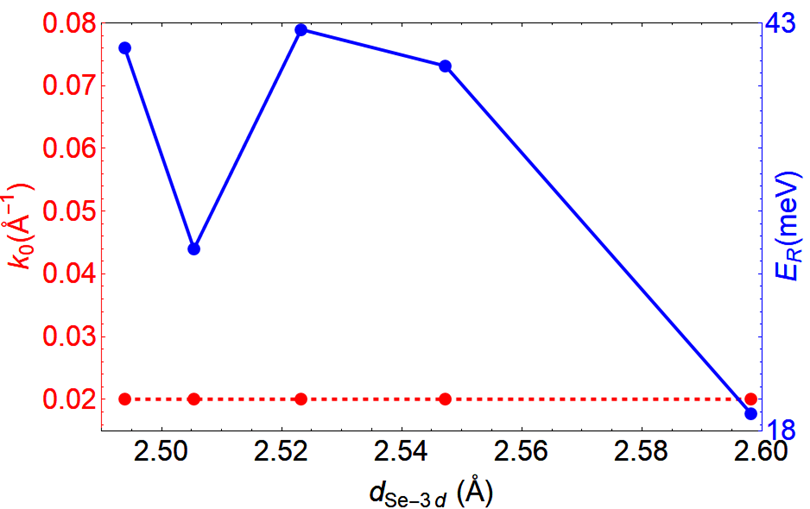}
\includegraphics[width=\columnwidth,angle=0]{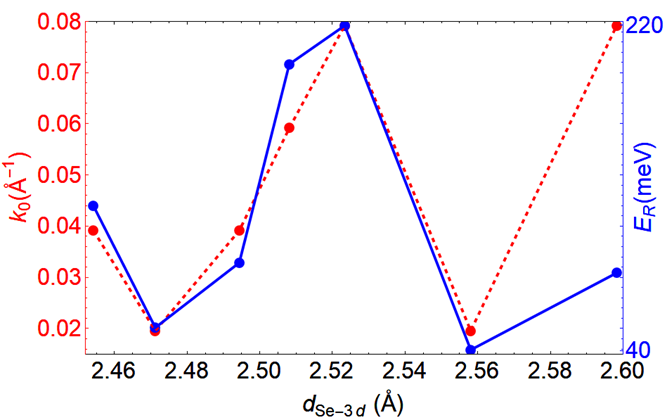}
\caption{Rashba as a function of the distance between the Se termination and the transition metal in the trial phase (top panel) and in the topological phase (bottom panel).}
\label{distance}
\end{figure}

In the asymmetric case, the two-fold degeneracy disappears in the presence of SOC and the Rashba effect rises in both trivial and topological phases. In the trivial phase, the Rashba effect is present in both the topmost valence and bottommost conduction band, as shown in the left panel of Fig. \ref{asymmetric}. The presence of the camelback features significantly reduces the band gap, since they form between the valence and conduction bands. Therefore, the gap is still direct, but it moves away from $\bar{\Gamma}$.
In the topological case, the giant Rashba effect appears. The system hosts the hybridization of the Rashba and Dirac-like. While in the trivial case, the topmost valence band and bottommost conduction bands were Rashba-like, in the topological case, they become Dirac-like, while the Rashba effect persists just in the cationic conduction band due to the larger SOC of the Pb. The two Dirac points are distinct in energy, as shown in the right panel of Fig. \ref{asymmetric}. From the eigenvectors, we confirm that the Dirac point above the Fermi level comes mainly from the Pb-surface, while the Dirac point below the Fermi level comes mainly from the Se-surface.\\
The Rashba effect vanishes at the $\bar{\Gamma}$ point, so the Dirac points remain unaffected there; however, minor deviations from the linear Dirac dispersion are observed away from $\bar{\Gamma}$ due to the Rashba effect.  
The system hosts Dirac surface states as long as the thickness of the thin film is sufficiently large to avoid the hybridization of the Dirac cones between the two surfaces. Indeed, in the case of hybridization between the Dirac points, the gap is open at the Fermi level.
In the present and all the following cases, it is always possible to observe flat bands, which are QLs, at the $\Gamma$ point in both the conduction and valence bands. They tend to become flatter the closer they are to the Fermi level and they were experimentally reported\cite{phdthesisBartek}.

\subsection{Slabs with covered surface: Rashba coefficient as a function of the distance between the metal and the topological insulator}

In this subsection, we introduce the metallic cover and focus solely on the topological phase. We examine the cases of the 3$d$ metal in the hollow position, which is the ground state according to DFT calculations. While experimentally this effect is observed for several 3d metals\cite{TUROWSKI2023155434} which are nonmagnetic for low thickness, within our calculations we need a non-magnetic 3d metal far from half filling, otherwise the metallic bands will cover the bands of the topological insulator, therefore, we opt for the Cu cover. More information is given in the Appendix.
We fixed the coordinates of the PbSe while we allowed the structural relaxation of the 3$d$ metal covering the surface.
The 3$d$ atoms remain on the surface without intercalation, with their relaxed z coordinate slightly above that of the uppermost Se layer. This arrangement forms a quasi-2D layer composed of the 3$d$ and Se atoms, with a Se–3$d$ distance of 2.45 {\AA}.
However, we observe a Rashba effect smaller than the uncovered surface, as shown in Fig. \ref{Bandstructure_3d_hollow}. We can tune this distance d$_{Se-3d}$=2.45 {\AA} by increasing it up to 2.60 {\AA}, analyzing the evolution of the Rashba properties in the trivial and topological phases. Our results are reported in Fig. \ref{distance}. We define the quantities k$_0$ and E$_R$ as in Figure \ref{coexistence}(a). We extract k$_0$ and E$_R$ from our DFT calculations. 

The quantities k$_0$ and E$_R$  are uncorrelated in the trivial as observed in the top panel of Figure \ref{distance}. While k$_0$ is constant, E$_R$ evolves in a non-monotonic way. In the trivial region, we obtain $\alpha_R$=4.0 eV{\AA} at the distance of 2.45 {\AA}, while it reduces by increasing the d$_{Se-3d}$. 
In the topological phase, the theoretical values of the parameters for the distance of 2.45 {\AA} are k$_0$=0.039\AA$^{-1}$, E$_R$=119.9 meV and $\alpha_R$=6.12 eV {\AA} where the experimental measure reaches the maximum value of $\alpha_R^{exp}$=3.51 eV{\AA}\cite{phdthesisBartek}. 
The evolution of the Rashba properties in the topological phase is reported in the bottom panel of Figure \ref{distance}. There is a correlation between these two properties that keeps the Rashba parameter almost constant. This correlation is ascribed to the interplay of the Dirac at the $\Gamma$ point and the Rashba.  
By tuning the distance d$_{Se-3d}$, the Rashba changes and we show here the dependence of the Rashba parameters on the distance of 3$d$ from the last layer of Se. This oscillating behavior in the topological regime could be attributed to the oscillation of the charge of the topological surface states\cite{Eremeev14}. Our theoretical results show that the Rashba coefficient is larger in the topological phase, confirming that the topological phase can produce a giant Rashba.

\subsection{Slabs with covered surface: Surface resolved band structure}

\begin{figure}[t!]
\centering
\includegraphics[height=\columnwidth, angle=270]{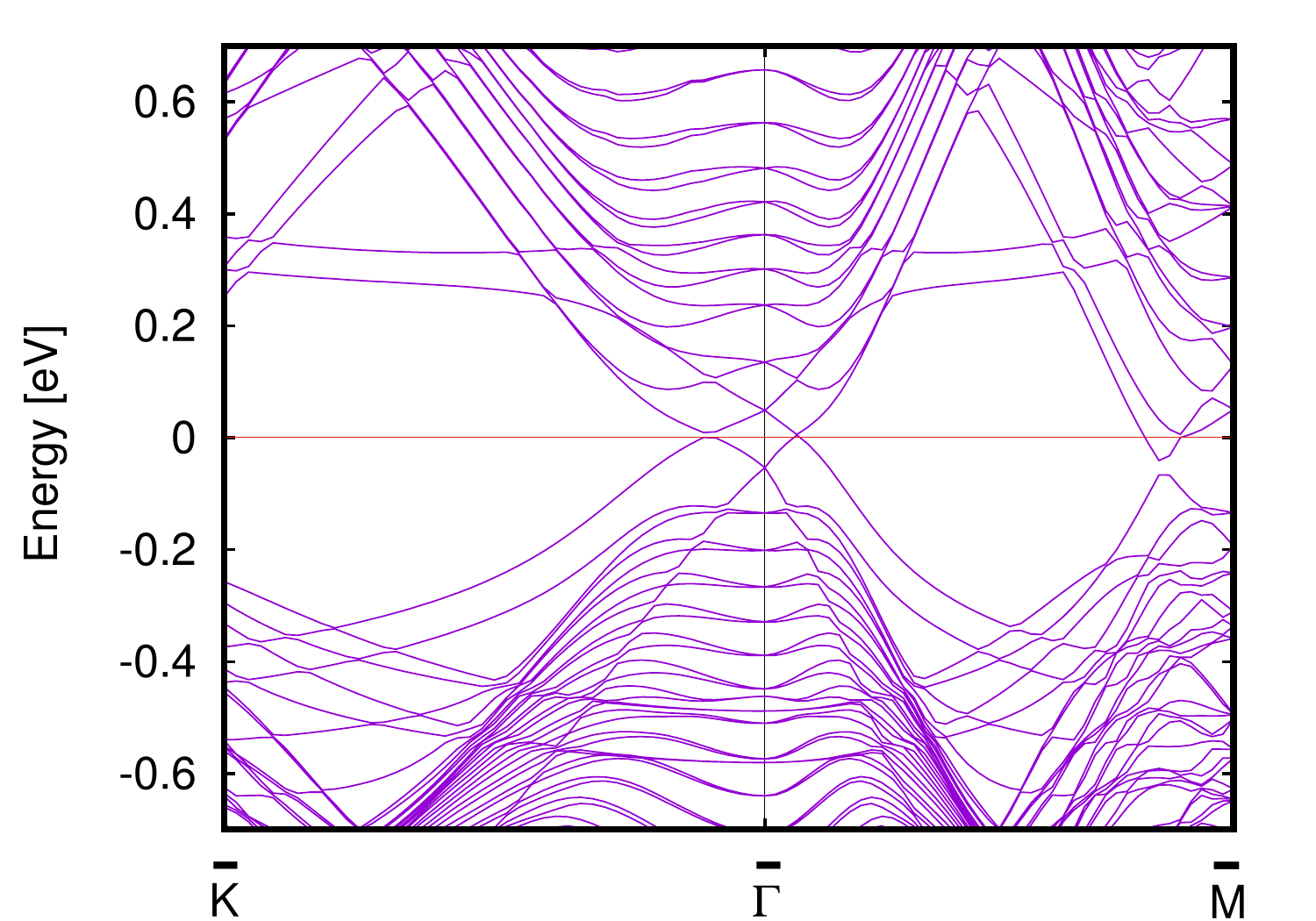}
\caption{Band structure of an asymmetric slab covered by the 3$d$ metal in the top position and in the topological phase. The Fermi level is set at zero energy.
}
\label{Bandstructure_3d_top_withoutU}
\end{figure}

DFT simulations fail to accurately determine the adatom position, favoring the hollow configuration\cite{Feibelman01}, while more advanced methods are required for a precise description
\cite{Stroppa08}. However, given the size of the system under investigation, we cannot go beyond the GGA+U exchange-correlation functional with the inclusion of the SOC interaction. Since the physical properties described in the experiments are better reproduced with the top position of the 3$d$ metal, we assume that the metallic layer is in the top position above the Se atoms of the surface. When the metal is in the top position, every 3$d$ atom is bonded to one Se atom, while when the metal is in a hollow position, every 3$d$ atom is bonded to three Se atoms. Therefore, the hollow position produces a large hybridization between the metal and the TCI. This results in greater hybridization between Dirac states from the PbSe surface and Rashba states within an extended region from the metal cover to the inner layers.

As in the asymmetric uncovered case, we have two Dirac points from two different surfaces, namely, the Se-Dirac point and the Pb-Dirac point. With the 3$d$ metal in the hollow position, we obtain that the two Dirac points at $\bar{\Gamma}$ come from the opposite surfaces with respect to the experimental observation.
Namely, the Dirac point of the topmost valence bands mainly comes from the top layers and, therefore, can be defined as the Se-Dirac point, while the Dirac point of the bottommost conduction bands mainly comes from the bottom layers and can be defined as the Pb-Dirac point. 
We also varied the distance of the 3$d$ metal from the last layer of Se and calculated the Rashba parameters' dependence on the distance 3$d$-Se.
We found that the points do not follow linear behavior.
To obtain the same nature of the Dirac points as in the experiments, we have to consider the 3$d$ metal in the top position.
The crystal structure of PbSe(111) with the full cover of 3$d$ atoms in the top position is reported in Fig. \ref{structure}(a).

\begin{figure}[t!]
\centering
\includegraphics[width=8.3cm]{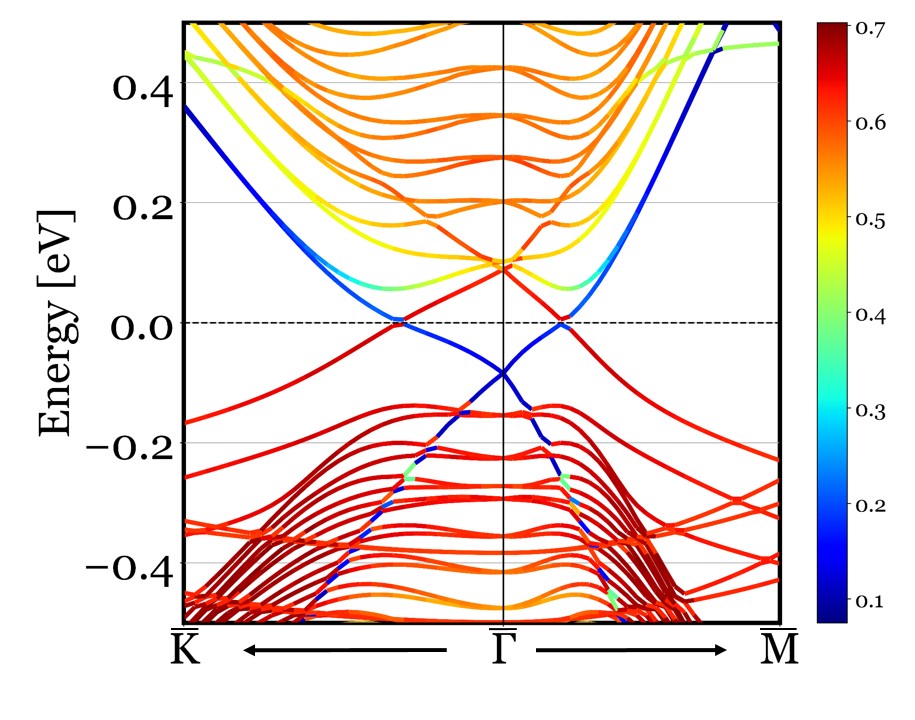}
\includegraphics[width=8.3cm]{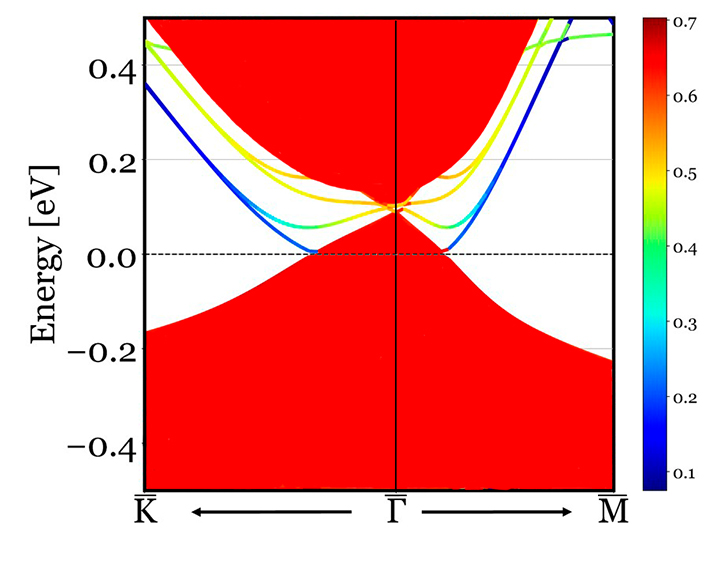}
\caption{(Top panel) Surface resolved band structure in the case of an asymmetric slab covered by the 3$d$ metal in the top position and the topological phase. The red color is used for the bands originating from the covered surface, while the blue is for the bands not originating from the covered surface. We have added a Coulomb repulsion U$_{s}$=10 eV on the s orbital of Cu and U$_{p}$=4 eV on the p orbitals of the last layer of Se. (Bottom panel) The same as in the top panel, with the contribution of the bulk filled in red. The Fermi level is set at zero energy. We plot in the energy range from -0.5 and 0.5 eV.
}
\label{surface_resolved}
\end{figure}

Without including Coulomb interaction, we obtain some trivial surface bands close to the Fermi level, as shown in Fig. \ref{Bandstructure_3d_top_withoutU}.
To move the trivial surface bands away from the Fermi level, we consider a Coulomb repulsion U$_{s}$=10 eV on the $s$-orbital of the metal and U$_{p}$=4 eV on the $p$-orbitals of the surface layer of Se. The metal in the top position, combined with the strong electronic correlations of the surface, pushes the Se-Dirac point above the Pb-Dirac point. This allows us to reproduce the experimental results as shown in the band structure reported in the top panel of Fig. \ref{surface_resolved}. In this case, the Rashba parameters of our band structure are k$_{0}$=0.049 {\AA}$^{-1}$ and E$_{R}$=33.9 meV, close to the experimental values which, for the considered case, are k$_{0}^{exp}$=0.031 {\AA}$^{-1}$ and E$_{R}^{exp}$=23.9 meV.\\

For a direct comparison with experimental ARPES results reported in Fig. \ref{coexistence}(d), we calculate the surface-resolved band structure by projecting the spectral weight of nearly all atomic layers, including the 3$d$ atomic layer, while excluding the bottom 12 layers (specifically, the lowest 6 of Pb and 6 of Se). The result is reported in the top panel of Fig. \ref{surface_resolved}.
We can see that we have two conduction bands outside the region of the bulk bands, in good agreement with experiments. One band has a Rashba-like feature with a minimum not in $\Gamma$, while the other band shows a flat region around the $\bar{\Gamma}$ point but with a minimum at the $\bar{\Gamma}$ point. 
In the bottom panel of Fig. \ref{surface_resolved} we report the same band structure of the top panel, but with the bulk bands colored in red, to better compare with the experimental result.
The blue bands are not visible in the experiment because they come from the bottom surfaces. 

We can conclude that the Rashba feature
mainly comes from the Pb bands distributed over a range of circa 8 nm before the surface. This allows a small hybridization with the Dirac point. The small hybridization together with the electronic correlation on the surface allows the superposition of the Dirac and Rashba points in covered Pb$_{1-x}$Sn$_x$Se.\\


\begin{figure}[t!]
\centering
\includegraphics[width=0.48\textwidth]{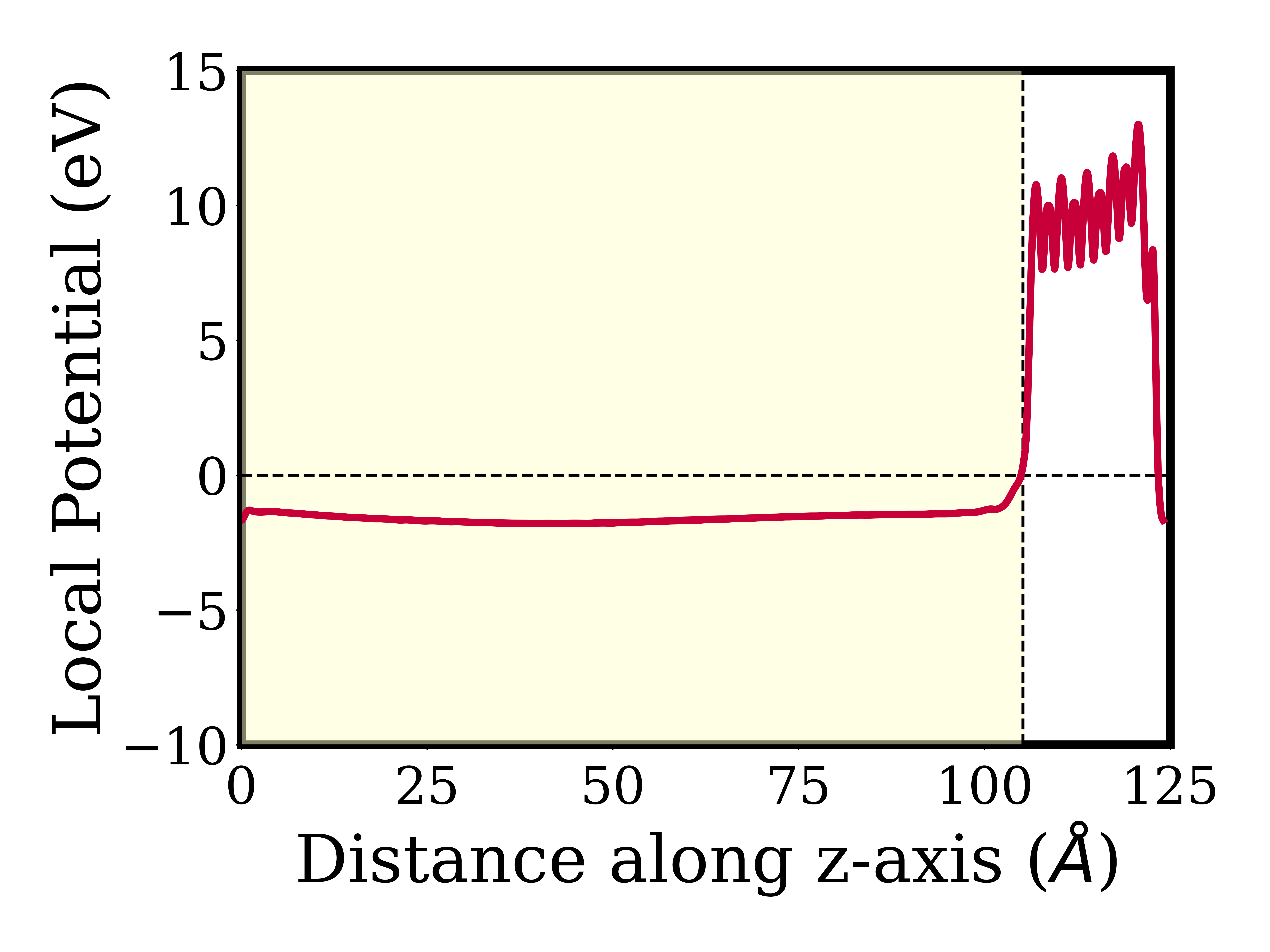}
\caption{$V_{surf}-V_{bulk}$ as a function of the distance along the z-axis. The yellow indicates the region of the TCI, while the white indicates the vacuum. The vertical dashed line indicates the surface layer of Se, on top of which there is the 3d cover.
}
\label{surf-bulk}
\end{figure}

\begin{figure*}[t!]
\centering
\includegraphics[width=0.48\textwidth]{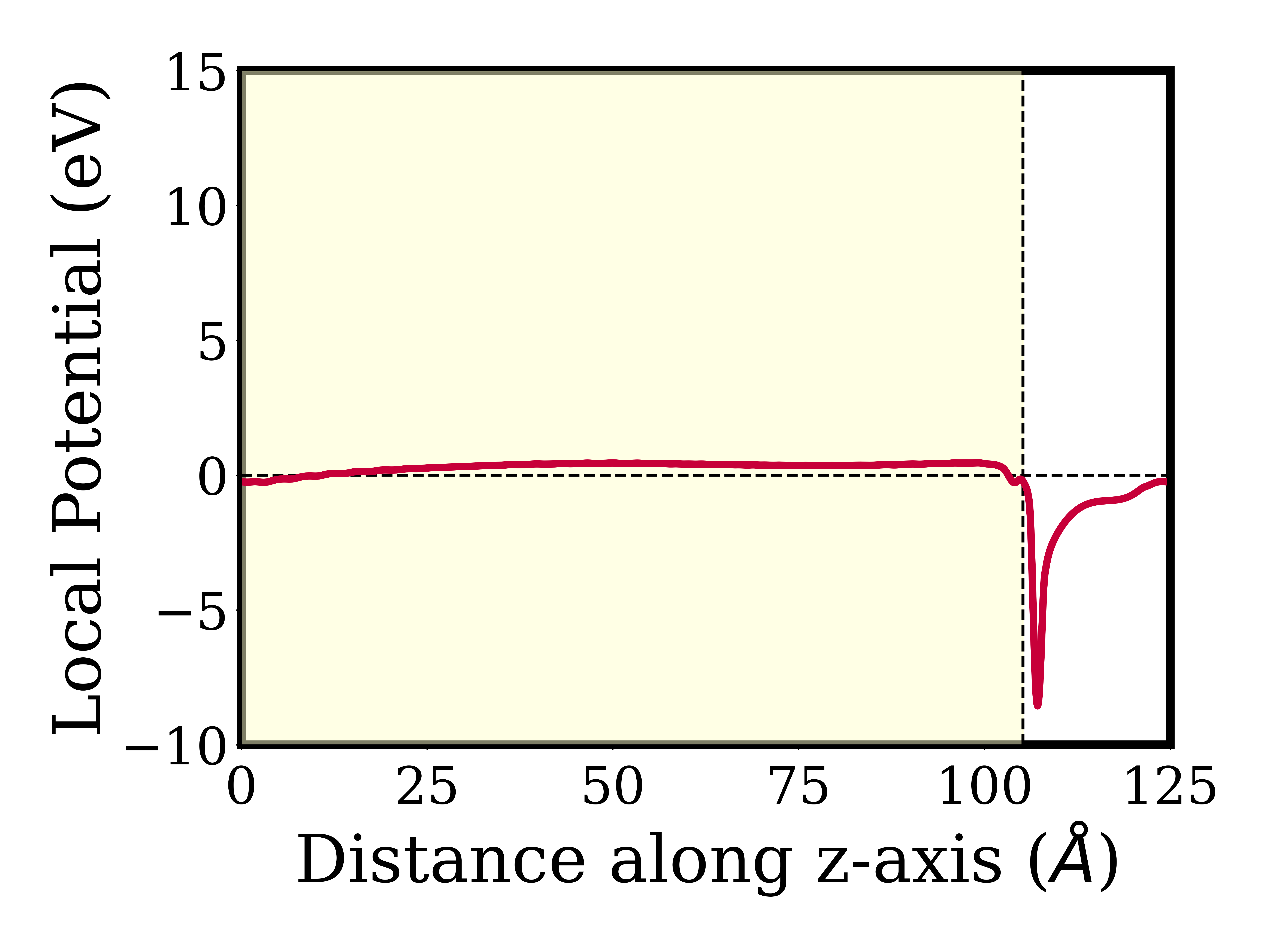}
\includegraphics[width=0.48\textwidth]{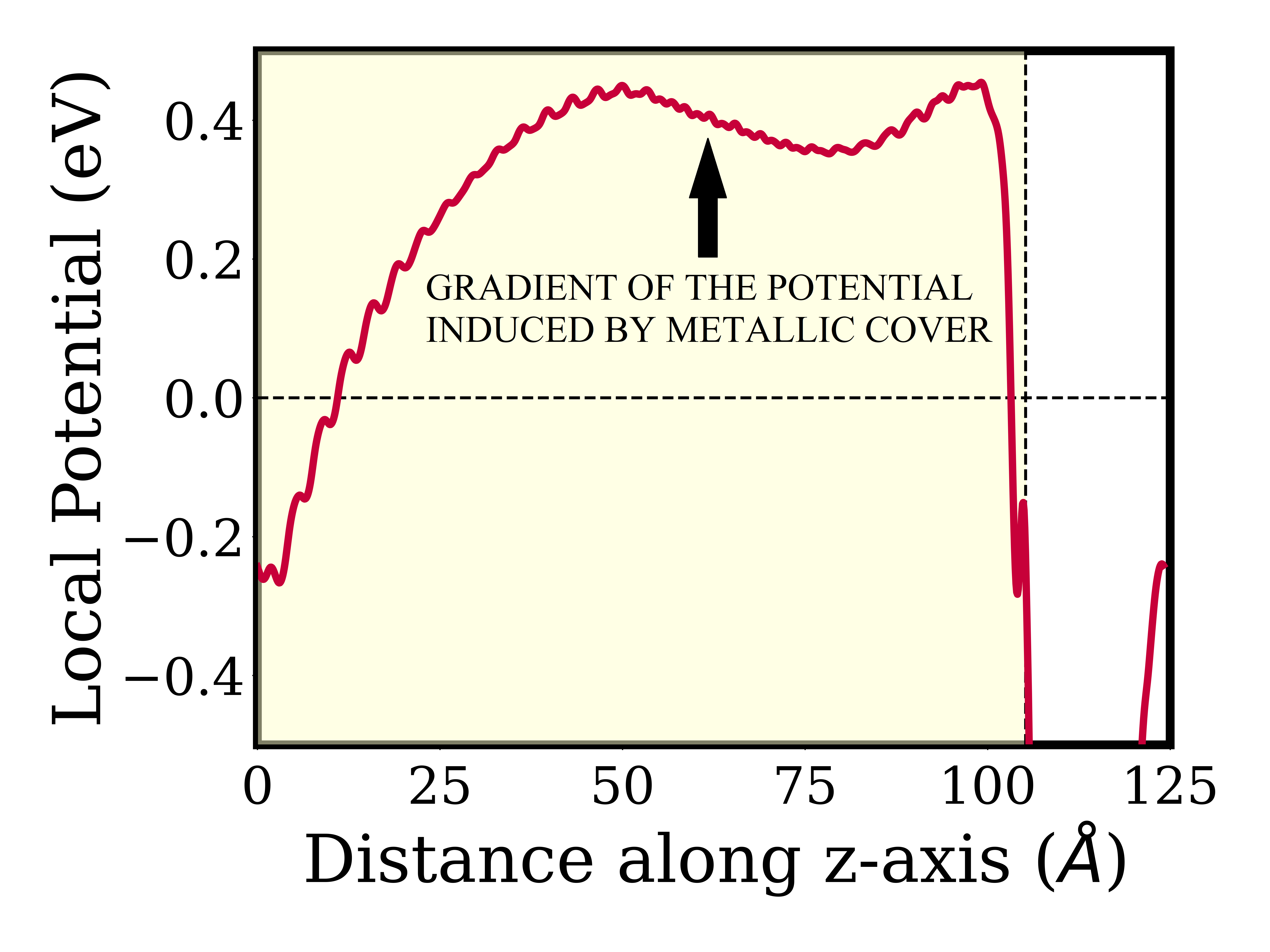}
\caption{(left panel) $V_{surf,3d}-V_{surf}$ as a function of the distance along the z-axis.
(right panel) Same as the left panel, but with the range of the potential between -0.5 and 0.5 eV. The yellow indicates the region of the TCI, while the white indicates the vacuum. The vertical dashed line indicates the surface layer of Se, on top of which there is the 3d cover. We can observe the gradient of the electrostatic potential, which will generate the Rashba bands, while the Dirac bands are on the surface.}
\label{surf3d-surf}
\end{figure*}

\subsection{Gradient of the electrostatic potential for the top position of the 3d metal}

PbSe is not ferroelectric but close to the ferroelectric transition\cite{10.1063/5.0222022}. We will show that the cover with the 3d element generates a gradient of the electrostatic potential, which is an electric field in the materials. This electric field is induced by the cover layer, even if it does not produce a full ferroelectric transition in the subsurface layer, but it definitely produces the Rashba effect. This electrostatic potential is crucial to reproducing the experimental results. We also calculated the effect of the SOC on the electrostatic potential in the case of a symmetric slab and we found that it is oscillating due to the presence of the Darwin term, with different contributions for ion and cation. However, in general, the potential generated by the SOC is relatively weak and it does not penetrate the materials. Therefore, the SOC interaction was not considered in the electrostatic potentials since the effect of SOC on the potential is relatively small. Once the Se-3d distance was fixed in the top position, we analyzed the effect of the 3$d$ metal on the electrostatic potential by calculating the difference between the potential in the case in which the 3$d$ covers the surface and the bulk case.
To have more insight, we define the quantities V$_{bulk}$, V$_{surf}$ and V$_{surf,3d}$ as the total electrostatic potential of the PbSe bulk, the PbSe uncovered surface and the PbSe surface covered with 3d, respectively. 

We are interested in the difference in the electrostatic potential of the covered surface with respect to the bulk:
\begin{equation*}
V_{surf,3d}-V_{bulk}=
(V_{surf,3d}-V_{surf})+   
(V_{surf}-V_{bulk})
\end{equation*}
that can be seen as the sum of the  following quantities:
\begin{eqnarray}
V_{3d}    &=&(V_{surf,3d}-V_{surf})\\   
V_{vacuum}&=&(V_{surf}-V_{bulk})
\end{eqnarray}

In Fig. \ref{surf-bulk}, we report the difference $V_{vacuum}=V_{surf}-V_{bulk}$ that represents the effect of the vacuum. As we can see, the electrostatic potential increases the energy of the PbSe surface, making the Dirac point not available in uncovered PbSe thin films. 
In the left panel of Fig. \ref{surf3d-surf}, we report the quantity $V_{surf,3d}-V_{surf}$ that describes the effect of the 3d coverage.
We observe that the 3$d$ electrostatic potential decreases the potential on the PbSe surface layer while increasing the potential on the inner layers. This band banding is in agreement with the experiments where the Fermi level of the surface is lower in energy than the Fermi level of the bulk, and the Fermi level when we cover with the 3$d$ metal is higher in energy than the Fermi level without the 3$d$. 
In the right panel of Fig. \ref{surf3d-surf}, we report the magnification of the left panel, We can observe a gradient of the electrostatic potential induced by the 3d cover along the samples. We propose that this gradient of the electrostatic potential is responsible for the Rashba bands, which are a few nm below the surface and are decoupled from the Dirac bands at the surface.

\section{Conclusions}

Investigating the case of the compressively strained PbSe(111), we demonstrated that the coexistence of pure Dirac and Rashba dispersions on the surface of the centrosymmetric topological insulator is only possible if the centrosymmetric topological insulator is non-homogeneous.
We demonstrate that a metallic overlayer induces a strong electrostatic potential gradient in the subsurface region in this polar system close to the ferroelectric transition\cite{10.1063/5.0222022}. 
This gradient of the electric field generates the electric field responsible for Rashba splitting in the subsurface layers. Consequently, the Rashba states arise from subsurface layers, while the Dirac states lie mainly on the surface layers. 

The results depend on the interplay of surface termination, the atomic position of the 3d cover and the electronic correlations on the surface.
In detail, to reproduce the experimental results, the 3d coverage should be in the top position. This is the only way to generate a strong gradient in the electrostatic potential, such as having the Dirac point from the Se surface above the Dirac point coming from the Pb surface. 
The Dirac point experimentally observed comes from the Se surface. The Rashba observed in the experiments mainly comes from the unoccupied bands, and, therefore, from the cationic layers not only close to the surface but also from the subsurface. 
Finally, we compare the properties of the Rashba in the trivial and topological phases. The topological phase enlarges the Rashba coefficient. The calculated Rashba coefficient agrees qualitatively with the experimental results. 
Our results open new directions for research into the synergy between broken inversion symmetry and topological phases.

\begin{acknowledgments}
We thank B. Turowski, T. Wojtowicz and V. V. Volobuev for describing their experimental results.
We acknowledge R. M. Sattigeri, T. Hyart and W. Brzezicki for their useful discussions.
This research was supported by the Foundation for Polish Science project “MagTop” no.~FENG.02.01-IP.05-0028/23, co-financed by the European Union from the funds of Priority 2 of the European Funds for a Smart Economy Program 2021–2027 (FENG). GC and CA acknowledge support from PNRR MUR project PE0000023-NQSTI.
We further acknowledge access to the computing facilities of the Interdisciplinary Center of Modeling at the University of Warsaw, Grant g91-1418, g91-1419, g96-1808 and g96-1809 for the availability of high-performance computing resources and support. We acknowledge the CINECA award under the ISCRA initiative IsC99 "SILENTS”, IsC105 "SILENTSG" and IsB26 "SHINY" grants for the availability of high-performance computing resources and support. We acknowledge the access to the computing facilities of the Poznan Supercomputing and Networking Center, Grants No. pl0267-01 and pl0365-01.
\end{acknowledgments}

\medskip

\appendix

\section{Computational details}

We performed density functional theory (DFT) calculations by using the VASP package \cite{Kresse93,Kresse96,Kresse96b}  based on the plane-wave basis set and the projector-augmented wave method \cite{Kresse99} with a cutoff of 268 eV for the plane wave basis. Relativistic effects, including spin-orbit coupling (SOC), were taken into account.
The calculations have been performed using 7$\times$7$\times$1 k-points centered in $\Gamma$ with 49 k-points in the independent Brillouin zone. 
We have optimized the internal degrees of freedom by
minimizing the total energy to be less than 1$\times10^{-4}$ eV.
As an exchange-correlation functional, the generalized gradient approximation (GGA) of Perdew, Burke, and Ernzerhof (PBE) has been adopted \cite{Perdew96}.
A vacuum spacer of $\sim$20 {\AA} was included to ensure negligible interaction between slabs.

We simulated the polar PbSe(111) surfaces considering both the symmetric and asymmetric cases. The symmetric case is when the upper and lower surfaces are identical (Pb terminated), instead in the asymmetric case, the lower surface is Pb terminated, while the upper surface is Se terminated. To avoid the polar catastrophe, the polar slab with alternating charges of +2$|e|$ and -2$|e|$ along the z-axis prefers to have the last layers with charges of +1$|e|$ and -1$|e|$. For this reason, we need to passivate the surfaces.
In the symmetric case, both the upper and lower surfaces were passivated by F atoms, while in the asymmetric case, the lower surface was passivated by F atoms, while the upper surface was passivated by H atoms.
The ideal Se and Pb surfaces produce strong dangling-bond surface states propagating in the energy gap over the entire 2D Brillouin zone, while the passivation of these surfaces with H or F atoms eradicates the trivial surface states\cite{Eremeev14}.

We have considered the PbSe(111) surface in the trivial and topological phases. To simulate the topological phase, we apply a hydrostatic 2\% compressive strain rather than Sn doping.
In the symmetric case, we considered 31 layers of Pb and 30 layers of Se in the topological phase and 29 layers of Pb and 28 layers of Se in the trivial phase. In the asymmetric case, we have considered 31 layers of Se and 31 layers of Pb in the topological phase, while 28 layers of Se and 28 layers of Pb in the trivial phase. We considered this large number of layers in the topological phase to avoid the wave functions of the upper and lower surface layers that would hybridize, opening a gap. Indeed, the hybridization between the Dirac points could produce an effect that is similar to the Rashba effect. Since we want to investigate the regime of pure Rashba effect, in our simulations, we always use a large number of layers.

The lattice constants in the unstrained trivial case are a=b=7.465 {\AA} and the distance between the Se atoms is d$_{Se-Se}$=$\frac{\sqrt{3}}{3}$a=4.310 {\AA}, while in the strained topological case are a=b=7.316 {\AA} and the distance between the Se atoms is d$_{Se-Se}$=$\frac{\sqrt{3}}{3}$a=4.224 {\AA}.

To reproduce the experimental coverage, we investigated the asymmetric case in which a 3$d$ metal covers the surface of Se, which is the experimental surface. We investigated both full and partial coverage. In the case of partial coverage, we obtain trivial surface states in the energy gap, and we do not reproduce the experimental results.
We cover with Cu to have Cu$^{+1}$ after charge transfer at the surface, Cu$^{+1}$ at the surface has d$^{10}$ electronic configuration that excludes overestimation of the magnetic properties as happens in monolayers in DFT. With respect to other results presented in the literature, we performed the relaxation procedure only for the layer occupied by the 3$d$ metal and we chose to freeze the other atomic positions to the bulk positions. We also investigated the cases in which we relax a few layers below the 3$d$ metal, but we obtained a Rashba effect that is one order of magnitude larger than the one experimentally observed\cite{Eremeev14}. This is because the surface relaxation done within DFT returns polar distortions much larger than those found in this material class\cite{Jin17}.
\\



\bibliography{PbSe111}
\end{document}